\documentstyle[12pt]{article}
\newlength{\extraspace}
\newlength{\extraspaces}
\newcommand{\be}{\begin{equation}
\addtolength{\abovedisplayskip}{\extraspaces}
\addtolength{\belowdisplayskip}{\extraspaces}
\addtolength{\abovedisplayshortskip}{\extraspace}
\addtolength{\belowdisplayshortskip}{\extraspace}}
\newcommand{\ee}{\end{equation}}
\newcommand{\ba}{\begin{eqnarray}
\addtolength{\abovedisplayskip}{\extraspaces}
\addtolength{\belowdisplayskip}{\extraspaces}
\addtolength{\abovedisplayshortskip}{\extraspace}
\addtolength{\belowdisplayshortskip}{\extraspace}}
\newcommand{\ea}{\end{eqnarray}}
\newcommand{\nonu}{\nonumber \\[.5mm]}
\newcommand{\A}{&\!\!\!}
\begin{document}
\thispagestyle{empty}
\begin{flushright}
SIT-LP-05/07 \\
{\tt hep-th/0507288} \\
July, 2005
\end{flushright}
\vspace{7mm}
\begin{center}
{\large{\bf On supersymmetry algebra \\[2mm]
based on a spinor-vector generator 
}} \\[20mm]
{\sc Kazunari Shima}
\footnote{
\tt e-mail: shima@sit.ac.jp} \ 
and \ 
{\sc Motomu Tsuda}
\footnote{
\tt e-mail: tsuda@sit.ac.jp} 
\\[5mm]
{\it Laboratory of Physics, 
Saitama Institute of Technology \\
Okabe-machi, Saitama 369-0293, Japan} \\[20mm]
\begin{abstract}
We study the unitary representation 
of supersymmetry (SUSY) algebra based on a spinor-vector generator 
for both massless and massive cases. 
A systematic linearization of nonlinear realization 
for the SUSY algebra is also discussed in the superspace formalism 
with a spinor-vector Grassmann coordinate. 
\end{abstract}
\end{center}

\newpage

Both linear (L) \cite{WZ} and nonlinear (NL) \cite{VA} 
supersymmetry (SUSY) are realized 
based on a SUSY algebra where spinor generators 
are introduced in addition to Poincar\'e generators. 
The relation between the L and the NL SUSY, 
i.e., the algebraic equivalence between various (renormalizable) 
spontaneously broken L supermultiplets and a NL SUSY action \cite{VA} 
in terms of a Nambu-Goldstone (NG) fermion 
has been investigated by many authors \cite{IK}-\cite{STT1}. 

An extension of the Volkov-Akulov (VA) model \cite{VA} of NL SUSY 
based on a spinor-vector generator, called the spin-3/2 SUSY, hitherto, 
and its NL realization in terms of a spin-3/2 NG fermion 
have been constructed by N.S. Baaklini \cite{Ba}. 
From the spin-3/2 NL SUSY model, 
L realizations of the spin-3/2 SUSY are suggested as corresponding 
supermultiplets to a spin-3/2 NL SUSY action \cite{Ba} 
through a linearization, although those have not yet known at all. 
Also the linearization of the spin-3/2 NL SUSY 
is useful from the viewpoint towards constructing 
a SUSY composite unified theory based on SO(10) super-Poincar\'e 
(SP) group (the superon-graviton model (SGM)) \cite{KS}. 
Indeed, it may give new insight into an analogous mechanism 
with the super-Higgs one \cite{DZ} 
for high spin fields which appear in SGM (up to spin-3 fields). 

In this letter, we investigate the unitary representation 
of the spin-3/2 SUSY algebra in \cite{Ba}
towards the linearization of the spin-3/2 NL SUSY. 
The role of the spinor-vector generator 
as creation and annihilation operators for helicity states 
is discussed explicitly. 
For both massless and massive representaions 
for the spin-3/2 SUSY algebra, 
we show examples of L supermutiplet-structure 
induced from those representations. 
We also discuss on a systematic linearization of the spin-3/2 NL SUSY 
in the superspace formalism with a spinor-vector Grassmann coordinate. 

Let us begin with the brief review of the spin-3/2 SUSY algebra 
and its NL realization in terms of the spin-3/2 NG fermion. 
The spin-3/2 SUSY algebra in \cite{Ba}, 
which satisfies all the Jacobi identities, 
is introduced based on a (Majorana) spinor-vector generator $Q^a_\alpha$ as 
\footnote{
Minkowski spacetime indices are denoted 
by $a, b, \cdots = 0, 1, 2, 3$ 
and four-component spinor indices are $\alpha, \beta, \cdots = (1), (2), (3), (4)$. 
The Minkowski spacetime metric is 
${1 \over 2}\{ \gamma^a, \gamma^b \} = \eta^{ab} = (+, -, -, -)$ 
and $\sigma^{ab} = {i \over 2}[\gamma^a, \gamma^b]$. 
We also use the spinor representation of the $\gamma$ matrices, 
$\gamma^0 = \pmatrix{ 0 & I \cr
                     I & 0 \cr
}$, 
$\gamma^i = \pmatrix{ 0 & - \sigma^i \cr
                      \sigma^i & 0 \cr
}$ with $\sigma^i$ being the Pauli matrices, 
$\gamma_5 = i \gamma^0 \gamma^1 \gamma^2 \gamma^3 
$ and the charge conjugation matrix, $C = - i \gamma^0 \gamma^2$. 
}
\ba
\A \A \{ Q^a_\alpha, Q^b_\beta \} 
= i \epsilon^{abcd} (\gamma_5 \gamma_c C)_{\alpha \beta} P_d, 
\label{spin3/2SUSY-QQ}
\\
\A \A [ Q^a_\alpha, P^b ] = 0, 
\label{spin3/2SUSY-QP}
\\
\A \A [ Q^a_\alpha, J^{bc} ] = {1 \over 2} (\sigma^{bc})_\alpha{}^\beta Q^a_\beta 
+ i (\eta^{ab} Q^c_\alpha - \eta^{ac} Q^b_\alpha), 
\label{spin3/2SUSY-QJ}
\ea
where $P^a$ and $J^{ab}$ are taranslational and Lorentz generators 
of the Poincar\'e group. 

The NL realization of the spin-3/2 SUSY which reflects Eq.(\ref{spin3/2SUSY-QQ}) 
is given by introducing the spin-3/2 (Majorana) NG fermion $\psi^a$ \cite{Ba}. 
Indeed, supertranslations of the $\psi^a$ and the Minkowski coordinate $x^a$ 
parametrized by a global (Majorana) spinor-vector parameter $\zeta^a$ 
are defined as 
\ba
\A \A 
\delta \psi^a = {1 \over \kappa} \zeta^a, 
\nonu
\A \A 
\delta x^a = \kappa \epsilon^{abcd} \bar\psi_b \gamma_5 \gamma_c \zeta_d, 
\label{stransln}
\ea
where $\kappa$ is a constant whose dimension is $({\rm mass})^{-2}$. 
Eq.(\ref{stransln}) means a NL SUSY transformation of $\psi^a$ 
at a fixed spacetime point, 
\be
\delta_Q \psi^a = {1 \over \kappa} \zeta^a 
- \kappa \epsilon^{bcde} \bar\psi_b \gamma_5 \gamma_c \zeta_d \partial_e \psi^a, 
\label{spin-3/2SUSY}
\ee
which gives the closed off-shell commutator algebra, 
\be
[ \delta_{Q1}, \delta_{Q2} ] = \delta_P(\Xi^a), 
\label{spin-3/2com}
\ee
where $\delta_P(\Xi^a)$ means a translation with a generator 
$\Xi^a = 2 \epsilon^{abcd} \bar\zeta_{1b} \gamma_5 \gamma_c \zeta_{2d}$. 

As parallel discussions in the VA model of the ordinary 
(spin-1/2) NL SUSY \cite{VA}, 
a spin-3/2 NL SUSY invariant differential one-form is given by 
\ba
\omega^a \A \A 
= d x^a + \kappa^2 \epsilon^{abcd} \bar\psi_b \gamma_5 \gamma_c d \psi_d 
\nonu
\A \A = (\ \delta^a_b 
+ \kappa^2 \epsilon^{acde} \bar\psi_c \gamma_5 \gamma_d \partial_b \psi_e\ ) \ dx^b 
\nonu
\A \A = (\ \delta^a_b + t^a{}_b\ ) \ dx^b 
\nonu
\A \A = w^a{}_b \ dx^b 
\label{one-form}
\ea
and also a NL SUSY invariant action constructed from Eq.(\ref{one-form}) 
has the form 
\ba
S = \A \A - {1 \over {2 \kappa^2}} 
\int \omega_0 \wedge \omega_1 \wedge \omega_2 \wedge \omega_3 
\nonu
= \A \A - {1 \over {2 \kappa^2}} 
\int d^4 x \ \vert w \vert 
\nonu
= \A \A 
- {1 \over {2 \kappa^2}} \int d^4 x 
\left[ 1 + t^a{}_a 
+ {1 \over 2}(t^a{}_a t^b{}_b - t^a{}_b t^b{}_a) \right. 
\nonu
\A \A 
\left. - {1 \over 3!} \epsilon_{abcd} \epsilon^{efgd} t^a{}_e t^b{}_f t^c{}_g 
- {1 \over 4!} \epsilon_{abcd} \epsilon^{efgh} t^a{}_e t^b{}_f t^c{}_g t^d{}_h 
\right]. 
\label{NLSUSYaction}
\ea
The second term in Eq.(\ref{NLSUSYaction}), i.e., $-(1/2 \kappa^2) t^a{}_a$, 
is the ordinary Rarita-Schwinger kinetic term for $\psi^a$. 

Since (spontaneously broken) spin-3/2 L supermultiplets are suggested 
from a linearization of the spin-3/2 NL SUSY, 
we investigate the structure of L supermultiplets 
induced from the spin-3/2 SUSY algebra 
(\ref{spin3/2SUSY-QQ}) to (\ref{spin3/2SUSY-QJ}). 
For this purpose, we first focus on the relation (\ref{spin3/2SUSY-QJ}) 
and discuss on the role of the spinor-vector generator $Q^a_\alpha$ 
as creation and annihilation operators for helicity states. 
When we choose the moving direction of a massless particle as the 3-axis, 
Eq.(\ref{spin3/2SUSY-QJ}) for the helicity operator $J^{12}$ becomes 
\ba
\A \A 
[ \ Q^0_\alpha, \ J^{12} \ ] = \pm {1 \over 2} Q^0_\alpha, 
\label{Q0J}
\\
\A \A 
[ \ Q^+_\alpha, \ J^{12} \ ] 
= {1 \over 2} Q^+_\alpha \ \ \ {\rm for} \ \alpha = (1), (3), 
\label{Q+J1/2}
\\
\A \A 
[ \ Q^+_\alpha, \ J^{12} \ ] 
= {3 \over 2} Q^+_\alpha \ \ \ {\rm for} \ \alpha = (2), (4), 
\label{Q+J3/2}
\\
\A \A 
[ \ Q^-_\alpha, \ J^{12} \ ] 
= - {3 \over 2} Q^-_\alpha \ \ \ {\rm for} \ \alpha = (1), (3), 
\label{Q-J3/2}
\\
\A \A 
[ \ Q^-_\alpha, \ J^{12} \ ] 
= - {1 \over 2} Q^-_\alpha \ \ \ {\rm for} \ \alpha = (2), (4), 
\label{Q-J1/2}
\\
\A \A 
[ \ Q^3_\alpha, \ J^{12} \ ] = \pm {1 \over 2} Q^3_\alpha, 
\label{Q3J}
\ea
where $Q^\pm_\alpha = (1/2) (Q^1_\alpha \pm i Q^2_\alpha)$. 
Eqs.(\ref{Q0J}), (\ref{Q+J1/2}), (\ref{Q-J1/2}) and (\ref{Q3J}) 
mean that the generators $Q^0_\alpha$, $Q^3_\alpha$, 
$Q^+_\alpha$ (for $\alpha = (1), (3)$) 
and $Q^-_\alpha$ (for $\alpha = (2), (4)$) 
raise or lower the helicity of states by 1/2 
as the same discussions in the spin-1/2 SUSY algebra. 
On the other hand, Eqs.(\ref{Q+J3/2}) and (\ref{Q-J3/2}) 
show that the operators $Q^+_\alpha$ (for $\alpha = (2), (4)$) 
and $Q^-_\alpha$ (for $\alpha = (1), (3)$) 
raise or lower the helicity of states by 3/2 
in contrast with the case of the spin-1/2 SUSY. 

Next from Eq.(\ref{spin3/2SUSY-QQ}) 
we study both massless and massive representations 
of the spin-3/2 SUSY algebra. 
Let us first discuss on the massless case, $P^2 = 0$. 
By choosing a light-like reference frame, 
where $P_a = (\epsilon, 0, 0, \epsilon)$, 
Eq.(\ref{spin3/2SUSY-QQ}) for $Q^0_\alpha$, $Q^\pm_\alpha$, 
$Q^3_\alpha$ becomes 
\ba
\A \A 
\{ Q^0_{(1)}, Q^+_{(3)} \} = \epsilon, \ \ \ \ \ 
\{ Q^0_{(2)}, Q^-_{(4)} \} = \epsilon, 
\nonu
\A \A 
\{ Q^0_{(3)}, Q^+_{(1)} \} = - \epsilon, \ \ \ \ \ 
\{ Q^0_{(4)}, Q^-_{(2)} \} = - \epsilon, 
\nonu
\A \A 
\{ Q^3_{(1)}, Q^+_{(3)} \} = - \epsilon, \ \ \ \ \ 
\{ Q^3_{(2)}, Q^-_{(4)} \} = - \epsilon, 
\nonu
\A \A 
\{ Q^3_{(3)}, Q^+_{(1)} \} = \epsilon, \ \ \ \ \ 
\{ Q^3_{(4)}, Q^-_{(2)} \} = \epsilon, 
\nonu
\A \A 
\{ Q^+_{(2)}, Q^-_{(3)} \} = \epsilon, \ \ \ \ \ 
\{ Q^-_{(2)}, Q^+_{(3)} \} = - \epsilon, 
\label{QQ-cr.an.}
\ea
and all other anticommutators vanish. 
Note that the 14 generators appear in Eq.(\ref{QQ-cr.an.}) 
(i.e., the two generators, $Q^+_{(4)}$ and $Q^-_{(1)}$, do not appear). 
In order to find an example of the generators in the Fock space, 
we divide Eq.(\ref{QQ-cr.an.}) into the following three (irreducible) parts, 
\ba
\A \A 
\{ Q^0_{(1)}, Q^+_{(3)} \} = \epsilon, \ \ \ \ \ 
\{ Q^0_{(4)}, Q^-_{(2)} \} = - \epsilon, 
\nonu
\A \A 
\{ Q^3_{(1)}, Q^+_{(3)} \} = - \epsilon, \ \ \ \ \ 
\{ Q^3_{(4)}, Q^-_{(2)} \} = \epsilon, 
\nonu
\A \A 
\{ Q^-_{(2)}, Q^+_{(3)} \} = - \epsilon, 
\label{QQ-cr.an.1}
\ea
and 
\ba
\A \A 
\{ Q^0_{(2)}, Q^-_{(4)} \} = \epsilon, \ \ \ \ \ 
\{ Q^0_{(3)}, Q^+_{(1)} \} = - \epsilon, 
\nonu
\A \A 
\{ Q^3_{(2)}, Q^-_{(4)} \} = - \epsilon, \ \ \ \ \ 
\{ Q^3_{(3)}, Q^+_{(1)} \} = \epsilon 
\label{QQ-cr.an.2}
\ea
for the generators 
which raise or lower the helicity of states by 1/2, 
while 
\be
\{ Q^+_{(2)}, Q^-_{(3)} \} = \epsilon 
\label{QQ-cr.an.3}
\ee
for the generators 
which raise or lower the helicity of states by 3/2. 
We further define creation and annihilation operators 
from appropriately rescaled generators as 
\ba
\A \A 
a_1 = 
{1 \over \sqrt{\epsilon}} 
(\ \xi_1 Q^0_{(1)} + \xi_2 Q^3_{(1)} 
+ \xi_3 Q^-_{(2)}\ ), 
\nonu
\A \A 
a^\dagger_1 = 
{1 \over \sqrt{\epsilon}} 
(\ - \xi_1 Q^0_{(4)} - \xi_2 Q^3_{(4)} 
+ \xi_3 Q^+_{(3)}\ ), 
\nonu
\A \A 
a_2 = 
{1 \over \sqrt{\epsilon}} 
(\ \xi'_1 Q^0_{(3)} + \xi'_2 Q^3_{(3)} 
+ \xi'_3 Q^-_{(4)} \ ), 
\nonu
\A \A 
a^\dagger_2 = 
{1 \over \sqrt{\epsilon}} 
(\ \xi'_1 Q^0_{(2)} + \xi'_2 Q^3_{(2)} 
- \xi'_3 Q^+_{(1)} \ ), 
\nonu
\A \A 
a_3 = {1 \over \sqrt{\epsilon}} Q^-_{(3)}, 
\nonu
\A \A 
a^\dagger_3 = {1 \over \sqrt{\epsilon}} Q^+_{(2)}, 
\label{aa+def}
\ea
which are consistent with the Majorana condition of $Q^a_\alpha$. 
\footnote{
In the two-component spinor formalism, 
the components, ($Q^a_{(1)}$, $Q^a_{(2)}$), correspond 
to an undotted spinor, while the ($Q^a_{(3)}$, $Q^a_{(4)}$) 
are expressed as a dotted spinor. 
Also the Majorana condition of $Q^a_\alpha$ 
means $(Q^a_{(1)})^\dagger = - Q^a_{(4)}$ 
and $(Q^a_{(2)})^\dagger = Q^a_{(3)}$ 
by the hermitian (complex) conjugation 
(for example, see \cite{WB}). 
}
In Eq.(\ref{aa+def}) $(\xi_1, \xi_2, \xi_3)$ and $(\xi'_1, \xi'_2, \xi'_3)$ 
are arbitrary parameters 
which can be chosen as $\{ a_i, a^\dagger_i \} = 1$ for $i = 1, 2$. 

Then Eqs.(\ref{QQ-cr.an.1}) to (\ref{QQ-cr.an.3}) become 
the following anticommutation relations, 
\ba
\A \A 
\{ a_1, a^\dagger_1 \} = \{ 2(\xi_1 - \xi_2) - \xi_3 \} \xi_3, 
\nonu
\A \A 
\{ a_2, a^\dagger_2 \} 
= 2 (\xi'_1 - \xi'_2) \xi'_3, 
\nonu
\A \A 
\{ a_i, a_j \} = 0, \ \ \ \{ a^\dagger_i, a^\dagger_j \} = 0, 
\label{aa+12}
\ea
where $i, j = 1, 2$, 
while 
\ba
\A \A 
\{ a_3, a^\dagger_3 \} = 1, 
\nonu
\A \A 
\{ a_3, a_3 \} = 0, \ \ \ \{ a^\dagger_3, a^\dagger_3 \} = 0. 
\label{aa+3}
\ea
Namely, Eqs.(\ref{aa+12}) and (\ref{aa+3}) 
are equivalent to Eq.(\ref{QQ-cr.an.}) under the definition (\ref{aa+def}), 
although the physical meaning and the mathematical structure 
of Eq.(\ref{aa+def}) are not known. 

If we choose the values of $(\xi_1, \xi_2, \xi_3)$ and $(\xi'_1, \xi'_2, \xi'_3)$ 
in Eq.(\ref{aa+12}) as $\{ a_i, a^\dagger_i \} = 1$ for $i = 1, 2$, 
the $(a_i, a^\dagger_i)$ mean the operators in the Fock space 
which raise or lower the helicity of states by 1/2. 
Also the $(a_3, a^\dagger_3)$ in Eq.(\ref{aa+3}) 
are the operators in the Fock space 
which raise or lower the helicity of states by 3/2. 
Therefore, a massless irreducible representation 
for the spin-3/2 SUSY algebra induced 
from Eqs.(\ref{aa+12}) and (\ref{aa+3}) is 
\be
[ \ \underline{1} \left( +{3 \over 2} \right), \underline{2} (+1), 
\underline{1} \left( +{1 \over 2} \right), 
\underline{1} (0), \underline{2} \left( -{1 \over 2} \right), 
\underline{1} (-1) \ ] 
+ [\ {\rm CPT\ conjugate}\ ]. 
\label{irrep}
\ee
In Eq.(\ref{irrep}) $\underline{n} (\lambda)$ means the number of states $n$ 
for the helicity $\lambda$. 

Let us second investigate the algebra (\ref{spin3/2SUSY-QQ}) 
for the massive case, $P^2 = m^2$. 
By taking a rest frame momentum to be $P_a = (m, 0, 0, 0)$, 
Eq.(\ref{spin3/2SUSY-QQ}) for $Q^\pm_\alpha$, 
$Q^3_\alpha$ becomes 
\ba
\A \A 
\{ Q^3_{(1)}, Q^+_{(3)} \} = - m, \ \ \ \ \ 
\{ Q^3_{(2)}, Q^-_{(4)} \} = - m, 
\nonu
\A \A 
\{ Q^3_{(3)}, Q^+_{(1)} \} = m, \ \ \ \ \ 
\{ Q^3_{(4)}, Q^-_{(2)} \} = m, 
\nonu
\A \A 
\{ Q^+_{(1)}, Q^-_{(4)} \} = - {1 \over 2} m, \ \ \ \ \ 
\{ Q^-_{(1)}, Q^+_{(4)} \} = {1 \over 2} m, 
\nonu
\A \A 
\{ Q^+_{(2)}, Q^-_{(3)} \} = {1 \over 2} m, \ \ \ \ \ 
\{ Q^-_{(2)}, Q^+_{(3)} \} = - {1 \over 2} m, 
\label{QQ-cr.an.II}
\ea
and all other anticommutators vanish. 
Note that in Eq(\ref{QQ-cr.an.II}) 
$Q^0_\alpha$ is completely decoupled from the algebra, 
while the anticommutator $\{ Q^+_{(1)}, Q^-_{(4)} \}$ 
does not vanish and the two generators, $Q^+_{(4)}$ and $Q^-_{(1)}$, 
appear in contrast with Eq.(\ref{QQ-cr.an.}). 

As in the massless case, 
we divide Eq.(\ref{QQ-cr.an.II}) into four (irreducible) parts as 
\ba
\A \A 
\{ Q^3_{(1)}, Q^+_{(3)} \} = - m, \ \ \ \ \ 
\{ Q^3_{(4)}, Q^-_{(2)} \} = m, 
\nonu
\A \A 
\{ Q^-_{(2)}, Q^+_{(3)} \} = - {1 \over 2} m, 
\label{QQ-cr.an.II1}
\ea
and 
\ba
\A \A 
\{ Q^3_{(2)}, Q^-_{(4)} \} = - m, \ \ \ \ \ 
\{ Q^3_{(3)}, Q^+_{(1)} \} = m, 
\nonu
\A \A 
\{ Q^+_{(1)}, Q^-_{(4)} \} = - {1 \over 2} m 
\label{QQ-cr.an.II2}
\ea
for the generators 
which raise or lower the helicity of states by 1/2, 
while 
\be
\{ Q^+_{(2)}, Q^-_{(3)} \} = {1 \over 2} m, 
\label{QQ-cr.an.II3}
\ee
and 
\be
\{ Q^-_{(1)}, Q^+_{(4)} \} = {1 \over 2} m, 
\label{QQ-cr.an.II4}
\ee
for the generators 
which raise or lower the helicity of states by 3/2. 
We also define creation and annihilation operators 
from appropriately rescaled generators as 
\ba
\A \A 
a_1 = {1 \over \sqrt{m}} (\ \eta_1 Q^3_{(1)} + \eta_2 Q^-_{(2)} \ ), 
\nonu
\A \A 
a^\dagger_1 = {1 \over \sqrt{m}} (\ - \eta_1 Q^3_{(4)} + \eta_2 Q^+_{(3)} \ ), 
\nonu
\A \A 
a_2 = {1 \over \sqrt{m}} (\ \eta'_1 Q^3_{(3)} + \eta'_2 Q^-_{(4)} \ ), 
\nonu
\A \A 
a^\dagger_2 = {1 \over \sqrt{m}} (\ \eta'_1 Q^3_{(2)} - \eta'_2 Q^+_{(1)} \ ), 
\nonu
\A \A 
a_3 = \sqrt{2 \over m} Q^-_{(3)}, 
\nonu
\A \A 
a^\dagger_3 = \sqrt{2 \over m} Q^+_{(2)}, 
\nonu
\A \A 
a_4 = - \sqrt{2 \over m} Q^-_{(1)}, 
\nonu
\A \A 
a^\dagger_4 = \sqrt{2 \over m} Q^+_{(4)}, 
\label{aa+defII}
\ea
which are consistent with the Majorana condition of $Q^a_\alpha$. 
In Eq.(\ref{aa+defII}) $(\eta_1, \eta_2)$ or $(\eta'_1, \eta'_2)$ 
are arbitrary parameters 
which can be chosen as $\{ a_i, a^\dagger_i \} = 1$ for $i = 1, 2$. 

Then Eqs.(\ref{QQ-cr.an.II1}) to (\ref{QQ-cr.an.II4}) become 
the following anticommutation relations, 
\ba
\A \A 
\{ a_1, a^\dagger_1 \} 
= - \left( 2 \eta_1 + {1 \over 2} \eta_2 \right) \eta_2, 
\nonu
\A \A 
\{ a_2, a^\dagger_2 \} 
= - \left( 2 \eta'_1 - {1 \over 2} \eta'_2 \right) \eta'_2, 
\nonu
\A \A 
\{ a_i, a_j \} = 0, \ \ \ \{ a^\dagger_i, a^\dagger_j \} = 0, 
\label{aa+II12}
\ea
where $i, j = 1, 2$, 
while 
\ba
\A \A 
\{ a_3, a^\dagger_3 \} = 1, 
\nonu
\A \A 
\{ a_4, a^\dagger_4 \} = - 1, 
\nonu
\A \A 
\{ a_i, a_j \} = 0, \ \ \ \{ a^\dagger_i, a^\dagger_j \} = 0, 
\label{aa+II34}
\ea
where $i, j = 3, 4$. 

For the massive case, 
we choose the value of $(\eta_1, \eta_2)$ in Eq.(\ref{aa+II12}) 
as $\{ a_1, a^\dagger_1 \} = 1$, 
while the value of $(\eta'_1, \eta'_2)$ as $\{ a_2, a^\dagger_2 \}$ 
gives the negative norm, i.e., $\{ a_2, a^\dagger_2 \} = - 1$. 
\footnote{
Athough the values of $(\eta_1, \eta_2)$ and $(\eta'_1, \eta'_2)$ 
can be chosen as $\{ a_i, a^\dagger_i \} = 1$ for $i = 1, 2$, 
the excess of helicity-$\pm1$ states 
in the resultant irreducible representation is not adequate 
for the massive case. 
}
Then the only $(a_1, a^\dagger_1)$ mean the operators in the Fock space 
which raise or lower the helicity of states by 1/2. 
Also the $(a_3, a^\dagger_3)$ in Eq.(\ref{aa+II34}) are 
the operators in the Fock space 
which raise or lower the helicity of states by 3/2, 
while the $\{ a_4, a^\dagger_4 \}$ in Eq.(\ref{aa+II34}) 
gives the negative norm. 
Therefore, a (physical) massive irreducible representation 
for the spin-3/2 SUSY algebra 
induced from $(a_1, a^\dagger_1)$ and $(a_3, a^\dagger_3)$ uniquely has 
the following structure, 
\be
[ \ \underline{1} \left( +{3 \over 2} \right), \underline{1} (+1), 
\underline{1} (0), \underline{1} \left( -{1 \over 2} \right) \ ] 
+ [\ {\rm CPT\ conjugate}\ ], 
\label{irrep2}
\ee
which represents massive spin-3/2, vector and scalar fields on shell. 

Explicit L realizations for the massless 
or the massive representations, e.g., Eqs.(\ref{irrep}) or (\ref{irrep2}), 
are now under investigation. 

Finally we comment on the systematic linearization of the spin-3/2 NL SUSY 
in the superspace formalism 
by introducing a spinor-vector Grassmann coordinate $\theta^a$. 
Indeed, let us denote a L superfield on the superspace coordinates 
$(x^a, \theta^a)$ by $\Phi(x^a, \theta^a)$, 
and define specific supertranslations as 
\ba
\A \A x'^a = x^a 
- \kappa \epsilon^{abcd} \bar\theta_b \gamma_5 \gamma_c \psi_d, 
\nonu
\A \A 
\theta'^a = \theta^a - \kappa \psi^a, 
\label{newcd}
\ea
which are just the spin-3/2 SUSY version of the specific supertranslations 
introduced in \cite{IK,UZ}. 
Then we can prove that the superfield on $(x'^a, \theta'^a)$, 
i.e., $\Phi(x'^a, \theta'^a) 
= \tilde \Phi(x^a, \theta^a; \psi^a(x))$, 
transforms homogeneously; namely, 
according to superspace translations of $(x^a, \theta^a)$ 
generated by the global (Majorana) spinor-vector parameter $\zeta^a$, 
\footnote{
The specific supertranslations (\ref{newcd}) 
correspond to the supertranslations (\ref{newstr}) 
by replacing the $\zeta^a$ with $- \kappa \psi^a$, 
while the NL SUSY transformation (\ref{spin-3/2SUSY}) 
(or Eq.(\ref{stransln})) is defined on a hypersuface 
$\theta^a = \kappa \psi^a$ in Eq.(\ref{newstr}). 
}
\ba
\A \A x'^a = x^a + \epsilon^{abcd} \bar\theta_b \gamma_5 \gamma_c \zeta_d, 
\nonu
\A \A 
\theta'^a = \theta^a + \zeta^a. 
\label{newstr}
\ea
in addition to the spin-3/2 NL SUSY transformation (\ref{spin-3/2SUSY}), 
the $\tilde \Phi$ transforms as 
\be
\delta_\zeta \tilde \Phi 
= \xi^a \partial_a \tilde \Phi, 
\label{tildePhi-tr}
\ee
where $\xi^a = \kappa \epsilon^{abcd} \bar\zeta_b \gamma_5 \gamma_c \psi_d$. 
Eq.(\ref{tildePhi-tr}) means that the components of $\tilde \Phi$ 
do not transform into each other as in the case 
of the spin-1/2 SUSY \cite{IK,UZ}, 
and so the following conditions, 
\be
{\rm components\ of}\ \tilde \Phi = {\rm constant}, 
\label{conditions}
\ee
may give SUSY invariant relations which connect a spin-3/2 L SUSY action, 
if it exists, 
with the spin-3/2 NL SUSY one (\ref{NLSUSYaction}). 

We summarize our results as follows. 
We have studied the unitary representation of the spin-3/2 SUSY algebra 
introduced in \cite{Ba} for both massless and massive cases. 
The role of the spinor-vector generator $Q^a_\alpha$ 
as creation and annihilaiton operators, 
which raise or lower the helicity of states by 1/2 or by 3/2, 
has been disscussed explicitly in Eqs. from (\ref{Q0J}) to (\ref{Q3J}). 
By defining creation and annihilation operators from appropriately 
rescaled generators as in Eqs.(\ref{aa+def}) or (\ref{aa+defII}), 
the structure of the L supermultiplets induced from the spin-3/2 
SUSY algebra has been shown as Eq.(\ref{irrep}) for the massless case 
or Eq.(\ref{irrep2}) for the massive case. 
We have also shown the systematic linearization method by introducing 
the specific supertranslations of Eq.(\ref{newcd}).

\newpage

%
\newcommand{\NP}[1]{{\it Nucl.\ Phys.\ }{\bf #1}}
\newcommand{\PL}[1]{{\it Phys.\ Lett.\ }{\bf #1}}
\newcommand{\CMP}[1]{{\it Commun.\ Math.\ Phys.\ }{\bf #1}}
\newcommand{\MPL}[1]{{\it Mod.\ Phys.\ Lett.\ }{\bf #1}}
\newcommand{\IJMP}[1]{{\it Int.\ J. Mod.\ Phys.\ }{\bf #1}}
\newcommand{\PR}[1]{{\it Phys.\ Rev.\ }{\bf #1}}
\newcommand{\PRL}[1]{{\it Phys.\ Rev.\ Lett.\ }{\bf #1}}
\newcommand{\PTP}[1]{{\it Prog.\ Theor.\ Phys.\ }{\bf #1}}
\newcommand{\PTPS}[1]{{\it Prog.\ Theor.\ Phys.\ Suppl.\ }{\bf #1}}
\newcommand{\AP}[1]{{\it Ann.\ Phys.\ }{\bf #1}}

\end{document}